\documentclass[reprint, pra,floatfix,superscriptaddress]{revtex4-1}
\usepackage{times}
\usepackage{amsmath,amsthm,amsfonts,amssymb,bm}
\usepackage{textcomp, color}
\usepackage{graphicx}

\begin{document}

\title{Geometric phase gates with adiabatic control in electron spin resonance}
\author{Hua Wu}
\affiliation{Department of Materials, Oxford University, Oxford, OX1 3PH, UK}

\author{Erik M. Gauger}
\email{erik.gauger@nus.edu.sg}
\affiliation{Department of Materials, Oxford University, Oxford, OX1 3PH, UK}
\affiliation{Centre for Quantum Technologies, National University of Singapore, 3 Science Drive 2, Singapore 117543}

\author{Richard E. George}
\affiliation{Department of Materials, Oxford University, Oxford, OX1 3PH, UK}

\author{Mikko M\"ott\"onen}
\affiliation{QCD Labs, COMP Centre of Excellence, Department of Applied Physics, Aalto University, P.O. Box 13500, FI-00076 AALTO, Finland}
\affiliation{Low Temperature Laboratory (OVLL), Aalto University, P.O. Box 13500, FI-00076 AALTO, Finland}

\author{Helge Riemann}
\author{Nikolai V. Abrosimov}
\affiliation{Leibniz-Institut f\"ur Kristallz\"uchtung, 12489 Berlin, Germany}

\author{Peter Becker}
\affiliation{PTB Braunschweig, 38116 Braunschweig, Germany}

\author{Hans-Joachim Pohl}
\affiliation{VITCON Projectconsult GmbH, 07743 Jena, Germany}

\author{Kohei M. Itoh}
\affiliation{School of Fundamental Science and Technology, Keio University, Yokohama, Japan}

\author{Mike L. W. Thewalt}
\affiliation{Department of Physics, Simon Fraser University, Burnaby, BC, Canada}

\author{John J. L. Morton}
\email{Present address: London Centre for Nanotechnology, University College London, 17-19 Gordon St, London WC1H 0AH; jjl.morton@ucl.ac.uk}
\affiliation{Department of Materials, Oxford University, Oxford, OX1 3PH, UK}
\date{\today}

\begin{abstract}
High-fidelity quantum operations are a key requirement for fault-tolerant quantum information processing. Manipulation of electron spins is usually achieved with time-dependent microwave fields. In contrast to the conventional dynamic approach, adiabatic geometric phase operations are expected to be less sensitive to certain kinds of noise and field inhomogeneities. Here, we introduce an adiabatic geometric phase gate for the electron spin. Benchmarking it against existing dynamic and non-adiabatic geometric gates through simulations and experiments, we show that it is indeed inherently robust against inhomogeneity in the applied microwave field strength. While only little advantage is offered over error-correcting composite pulses for modest inhomogeneities $\lesssim 10 \%$, the adiabatic approach reveals its potential for situations where field inhomogeneities are unavoidably large.
\end{abstract}

\maketitle

\section{Introduction}

Precise coherent control of quantum systems is an essential ingredient for many quantum technologies. In particular, high-fidelity gate operations on quantum bits (qubits) are central to practical realizations of quantum information processing~\cite{Nielsen2000}. The electron spin provides a quantum two-level system that is well suited for the physical implementation of a qubit. Dynamic control of electron spin states is commonly realized by applying microwave pulses in electron spin resonance (ESR)~\cite{Schweiger2001}. Although single microwave pulses in conventional ESR spectrometers usually have non-negligible errors in amplitude and phase~\cite{Morton2005a}, high-fidelity single-qubit operations can often still be realised using carefully designed pulse sequences such as broadband composite pulses (BB1) ~\cite{Wimperis1994, Morton2005} and Knill pulses~\cite{Ryan2010, Souza2011}. 

A different approach to qubit operations involves geometric manipulations of the quantum system~\cite{Berry1984,Wilczek1984,Aharonov1987, Zanardi1999, Zhu2003}. This geometric approach to quantum computing is argued to be more robust against noise in the control parameters~\cite{Ekert2000, Zhu2005, Oreshkov2009, Solinas2010}. Geometric single- and two-qubit logic gates have been demonstrated in some systems such as nuclear spins~\cite{Suter1987,Jones2000,Du2003}, trapped-ions~\cite{Leibfried2003} and superconducting qubits~\cite{Leek2007,Mottonen2008}. For spin $1/2$ systems, theoretical calculations predict robustness against fluctuations in the static field and inhomogeneities in the microwave field~\cite{DeChiara2003};  this has also been explored experimentally with trapped ultracold neutrons~\cite{Filipp2009}. 

Here, we demonstrate the implementation of a single-qubit geometric phase gate using adiabatic control of electron spins. We show that under current experimental conditions  this leads to a much higher fidelity than achieved with simple dynamic phase gates. Interestingly, we find that the adiabatically obtained fidelity is comparable to that achieved by composite non-adiabatic pulses. These results are also verified by simulations, which indicate that the fidelity of the adiabatic geometric phase gate remains high when the inhomogeneity in the microwave field strength becomes large, unlike for the non-adiabatic approach. 

\section{Geometric phase}

After a cyclic evolution, a quantum system acquires a phase that depends on the geometric property of the evolution, the so-called Berry phase~\cite{Berry1984}. For an electron spin $1/2$, this geometric phase is determined by its trajectory on the Bloch sphere. Consider a spin initialised in the eigenstate $|0\rangle$ with respect to a static magnetic field along the $z$ axis, and with an additionally applied microwave field detuned from resonance [see Fig.~\ref{sequence}(a)]. Then, slowly tuning  the microwave frequency into resonance induces the eigenstate $|0(t)\rangle$ to adiabatically follow the effective magnetic field $B(t)$ in the rotating frame, rotating it into the $xy$ plane. The phase of the microwave drive can be swept to rotate the eigenstate by some angle $\phi$. Finally, the microwave field is detuned again, taking the eigenstate back to the $z$ axis. The geometric phase $\gamma_{|0\rangle}$ acquired by the state $|0\rangle$ is given by the enclosed solid angle  $\Theta$ of its trajectory on the Bloch sphere, $\gamma_{|0\rangle}=\Theta/2=\phi/2$. By the same analysis, the geometric phase acquired for the $|1\rangle$ state is $\gamma_{|1\rangle}= - \phi/2$, yielding a geometric phase for a general state of $\gamma=\gamma_{|0\rangle} - \gamma_{|1\rangle}= \phi$. 

In addition to the geometric phase, the electron spin also acquires a dynamic phase during the evolution given by $\delta = \int_0^{\tau} \frac{1}{\hbar}g\mu B(t)  S {\rm d}t$, where $\tau$ is the total length of the control sequence and $S$ is the expectation value of the spin operator, so that $ g \mu B(t)  S = E(t)$ is the time-dependent energy of the eigenstate. To remove this dynamic phase from the final state, we introduce a $\pi$ phase shift of the microwave field exactly halfway through the control sequence. The dynamic phase accumulated during the second half of the control sequence is $\delta_2=\int_{\frac{\tau}{2}}^{\tau} \frac{1}{\hbar}g\mu B(t)  S {\rm d}t = - \int_0^{\frac{\tau}{2}} \frac{1}{\hbar}g\mu B(t) S {\rm d}t = - \delta_1$, thus resulting in a vanishing final dynamic phase. We expect our adiabatic geometric phase gate to be particularly well suited for ESR and other schemes employing the rotating frame since the phase degree of freedom can be controlled with high speed and accuracy, providing a convenient cancellation of the dynamic phase.

\section{Model}

There are two options for tuning the initially off-resonant microwave field into resonance: first, by adding an offset to the static field $B_0$ in $z$ direction whose magnitude decreases in time. Second, by tuning the frequency of the microwave field. As the length of the adiabatic process is within a few microseconds, we employ the latter approach. 

\begin{figure}[t]
\begin{center}
\includegraphics[width=3.5in]{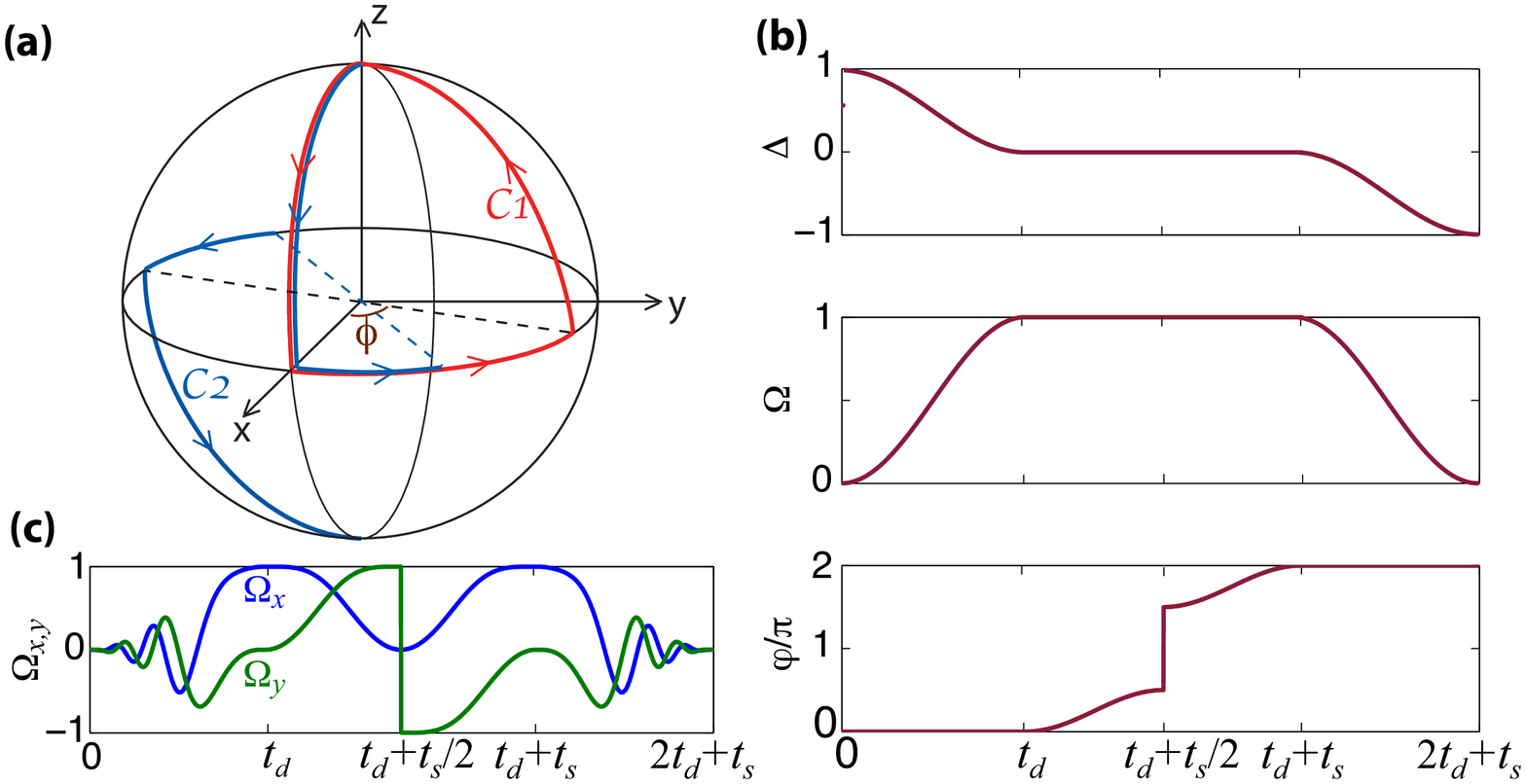}
\caption{(Color online) \textbf{(a)}: Evolution of the spin eigenstate $|0\rangle$ (red trace $\mathcal{C}_1$) represented by the spin vector $S$ and effective magnetic field arising from the applied microwave field (blue trace $\mathcal{C}_2$). \textbf{(b)}: The detuning $\Delta$, amplitude $\Omega$, and phase $\varphi$ of the microwave field during an adiabatic $\pi$ phase gate ($\Delta$ and $\Omega$ are shown in arbitrary units). The microwave frequency is tuned from off-resonance to resonance between time 0 and $t_d$ and then kept resonant for a period of $t_s$ while the phase $\phi$ of the microwave is swept. After time $t_d+t_s$ the microwave is detuned again. The effective Hamiltonian changes sign after the $\pi$ phase shift at $t_d+t_s/2$. \textbf{(c)}: The $x$ (blue) and $y$ (green) components of the microwave field applied to the electron spin for a $\pi$ phase gate (in arbitrary units). }\label{sequence}
\end{center}
\end{figure}

For a time-dependent microwave frequency $\omega(t)$, the transformation of the Hamiltonian from the laboratory frame to a rotating frame with subsequent rotating wave approximation (RWA) can be performed in several ways. For instance, one can use the canonical rotating frame with constant frequency $\omega_{\rm R}$, where the spin resonance frequency $\omega_0$ would be a natural choice for $\omega_{\rm R}$. However, the resulting Hamiltonian then features important time-dependent oscillatory terms. Alternatively, we can choose a rotating frame which always tracks the frequency of the driving microwave field $\omega_{\rm R}=\omega(t)$. In this case, the transformed Hamiltonian after the RWA is given in the eigenbasis of $S_z$ by
\begin{equation}
H = \hbar \begin{pmatrix}
\frac{1}{2}(\Delta + t \dot{\Delta}) & \Omega e^{- i\varphi}  \\
\Omega e^{i\varphi} & - \frac{1}{2}(\Delta + t \dot{\Delta})
\end{pmatrix},	\label{eq_ch6_Hr_time_dependent}
\end{equation}
where $\Omega, \varphi$ are the time-dependent amplitude and phase of the microwave field, $\Delta=\omega(t) - \omega_0$ is the detuning, and $\dot{\Delta}=\dot{\omega}(t)$ its time derivative.  For our simulations this form of the transformation is used, ensuring that the measurement of the spin magnetization is carried out in the same reference frame that is used for the depiction of the trajectory on the Bloch sphere in Fig.~\ref{sequence}(a).

Figures~\ref{sequence}(b) and (c) show the microwave profile used in this study for implementing an adiabatic geometric $\pi$ phase gate (see the Appendix \ref{appendix} for its mathematical parameterisation). The fast oscillations at the beginning and the end of the control sequence in Fig.~\ref{sequence}(c) arise due to the finite detuning, getting slower as the microwave field approaches resonance ($\Delta \to 0$), and disappearing entirely for $t_d\le t \le t_d+t_s$. The variation of the $x$ and $y$ components of the microwave field $\Omega_x$ and $\Omega_y$, respectively, during $t_d\le t \le t_d+t_s$ is due to the phase sweep of the microwave which causes the rotation of the spin magnetization in the $xy$-plane. 

Different choices for the microwave profile could be made, but vitally the time derivatives ($\dot{\Delta}, \dot{\Omega}, \dot{\varphi}$) must be kept continuous (except at time $t=t_d+t_s /2$) for achieving adiabatic evolution. However, continuity of the aforementioned parameters is not sufficient for meeting the adiabaticity condition, which also requires that the control sequence $\tau$ should take much longer than the spin precession, i.e.~$\tau\gg\Delta_0^{-1}$, where $\Delta_0$ is the initial detuning.  Experiment and simulation both confirm that with all the other parameters fixed, the fidelity of the adiabatic phase gate increases with $\tau$. 

\section{Simulation}

We simulate such an adiabatic $\pi$ phase gate and compare its performance with phase gates based on dynamic pulses. The adiabatic implementation is based on the pulse sequence shown in Fig.~\ref{sequence}(c), whereas the sequence for applying a dynamic phase $\phi$ is $(\frac{\pi}{2})_x  (\phi)_y  (\frac{\pi}{2})_{-x}$. The corresponding BB1 composite pulse sequence is built by replacing each single pulse with a composite pulse $(\theta)_{\varphi_0}(\pi)_{\varphi_1+\varphi_0}(2\pi)_{\varphi_2+\varphi_0}(\pi)_{\varphi_1+\varphi_0}$, where $\theta$ and $\varphi_0 \in \{ x, y, -x\}$ are the rotation angle and phase of the corresponding single pulse, respectively, and $\varphi_1= {\rm arccos}( - \theta/4\pi), \varphi_2=3\varphi_1$. It is also possible to construct a non-adiabatic geometric phase gate by applying two $\pi$ pulses of different phase successively $(\pi)_{\varphi_1}  (\pi)_{\varphi_2}$, and the BB1 version of this sequence can be built accordingly. The geometric phase acquired by this operation is determined by the phase difference between the two $\pi$ pulses: $\gamma=2(\varphi_2 - \varphi_1)$. 

\begin{figure}[t]
\begin{center}
\includegraphics[width=3.5in]{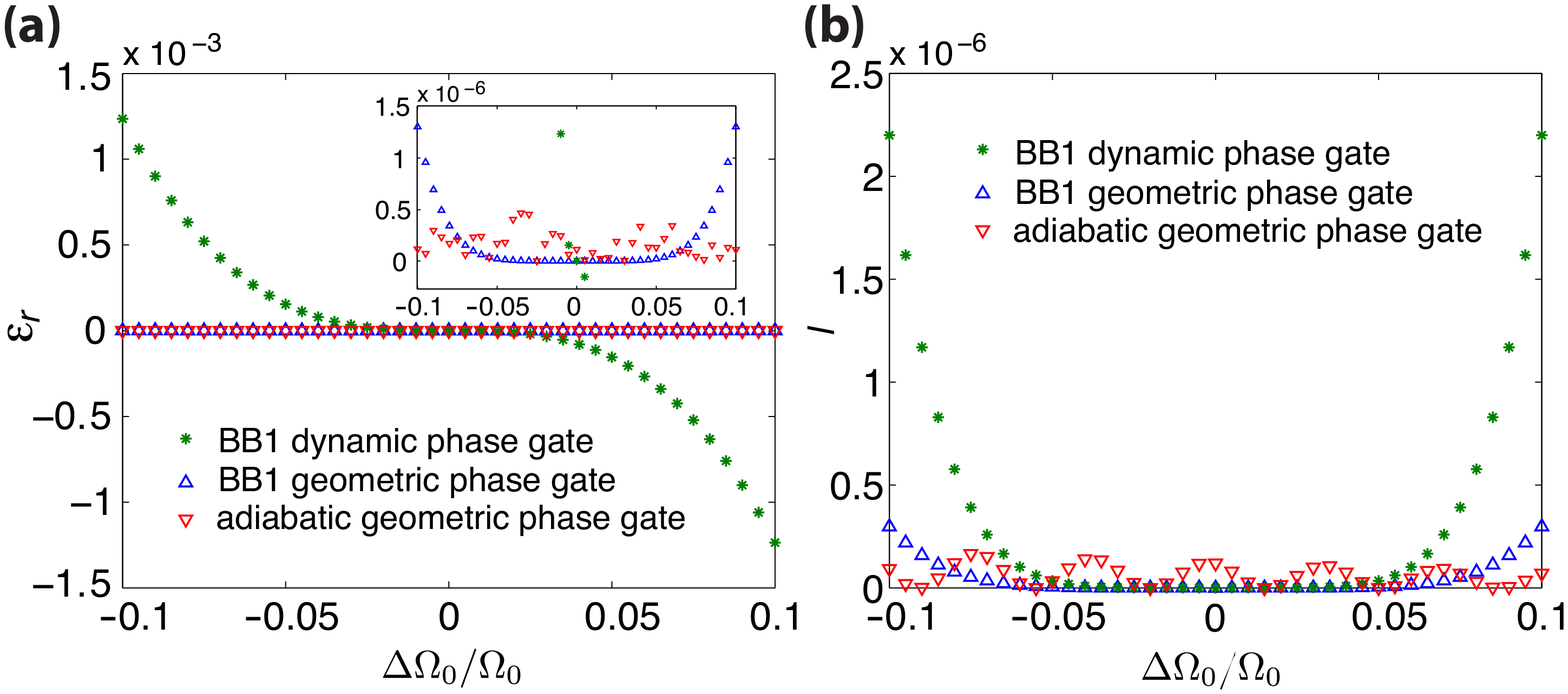}
\caption{(Color online) \textbf{(a)}: Simulated relative phase errors $\varepsilon_r=\gamma/\pi - 1$ for $\pi$ phase gates as functions of the error in microwave amplitude $\Delta\Omega_0/\Omega_0$, where $\gamma$ is the acquired phase in the simulation. The inset shows a zoomed in view of $\varepsilon_r$ for a better comparison of the two geometric gates. \textbf{(b)}: Infidelities $I$ (see text) of the operations as functions of $\Delta\Omega_0/\Omega_0$. All simulations employ a single spin $1/2$ and do not include inhomogeneous
broadening of the spin packet. The parameters are $\Omega_0= 14$~MHz, initial detuning $\Delta_0=6$~MHz, $t_d=2~\mu$s, $t_s=4~\mu$s. Note that since there is no microwave noise, the BB1 errors are independent of the BB1 pulse duration. }
\label{simulation}
\end{center}
\end{figure}

The simulated phase errors of the three different phase gates are shown as a function of the deviation from a given microwave amplitude in Fig.~\ref{simulation}(a). The maximum relative deviation $\Delta \Omega_0$ from its center value  $\Omega_0$ is set to $\pm 10\%$, a common value for typical ESR spectrometers. The results show that the phase error in the BB1 dynamic phase gate is much larger than for both geometric phase gates. The relative phase error has a third order dependence on the error in microwave amplitude for the BB1 dynamic phase gate, i.e.~$\varepsilon_{r,{\rm dyn}}\sim O\left[(\Delta\Omega_0/\Omega_0)^3\right]$, whereas it has a sixth order dependence for the BB1 geometric phase gate, i.e.~$\varepsilon_{r,{\rm geo}}\sim O\left[(\Delta\Omega_0/\Omega_0)^6\right]$. The relative phase error in the adiabatic geometric phase gate is comparable with the BB1 geometric phase gate in this range of microwave inhomogeneity. However, we can see that if the microwave field strength inhomogeneity becomes larger than 10\%, which is the case for most coplanar cavities, the performance of the adiabatic phase gate exceeds that of the BB1 geometric phase gate. Small fluctuations of the adiabatic gate are due to imperfect adiabaticity of the operation, decreasing both for a stronger microwave field or a slower passage.

The quality of a phase gate can also be characterised by the infidelity $I$ of the operation, defined as
\begin{equation}
I = 1 - \frac{|{\rm Tr}[UU_0^{-1}]|}{2} ,	\label{ad_infidelity}
\end{equation}
where $U_0$ and $U$ are the operators for an idealized and simulated $\pi$ phase gate, respectively. Figure~\ref{simulation}(b) shows the infidelities of the three phase gates as  functions of $\Delta\Omega_0/\Omega_0$. For the two phase gates using BB1 composite pulses, the infidelity increases as $O\left[(\Delta\Omega_0/\Omega_0)^6\right]$. The infidelity of the BB1 dynamic phase gate is much smaller than its phase error, because in the measurement of infidelity only the diagonal elements of $U$ are considered while the phase error $\varepsilon_r$ is more directly related to the off-diagonal elements of $U$. The oscillations in the infidelity of the adiabatic phase gate are due to the finite non-adiabaticity of the operation, but only have a minor effect on the geometric phase. The infidelity of the BB1 geometric  gate is comparable with the adiabatic variant within $\pm 10\%$ microwave inhomogeneity, and the adiabatic phase gate is more robust for larger inhomogeneity.

\section{Experiment}

\begin{figure}[t]
\begin{center}
\includegraphics[width=3.5in]{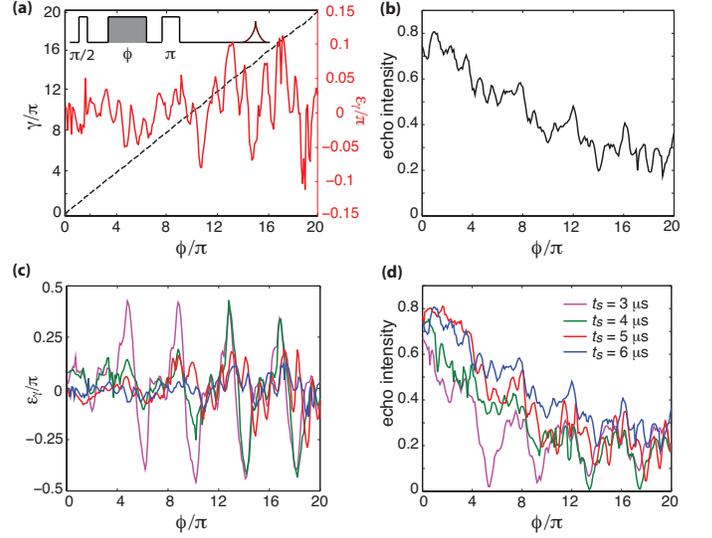}
\caption{(Color online) \textbf{(a)}: Measured geometric phase $\gamma$ (black, dashed) and error $\varepsilon_\gamma$ (red, solid) as functions of the angle $\phi$ that the spin rotates in the $xy$ plane during the adiabatic sequence. Inset: Pulse sequence for measuring the geometric phase gate. \textbf{(b)}: Intensities of the measured spin echo signal normalized to corresponding Hahn echo intensities. \textbf{(c)}: Phase error and \textbf{(d)}: Echo intensity for adiabatic phase gates of different $t_s$. In panel (c) $t_s$ decreases from bottom to top at $\phi = 5 \times \pi$, whereas in (d) it increases  from bottom to top at $\phi = 6 \times \pi$. In these experiments $t_d=2~\mu$s, maximum of the microwave amplitude $\Omega_0=0.64$~MHz, and initial detuning $\Delta_0=6$~MHz. In panel (a), $t_s = 6~{\rm \mu}$s. }
\label{ad_demonstration}
\end{center}
\end{figure}

For the experiments, we used a sample with narrow ESR linewidth (P donors in high-purity $^{28}$Si crystal at 8K) in order to ensure that all the spins are within the bandwidth of the adiabatic control sequence. The X-band microwave signal is generated at a constant frequency, which is then modulated by the $I/Q$ signals from an arbitrary waveform generator to create the required microwave field, such as the one shown in Fig.~\ref{sequence}(c). The complete sequence for measuring the geometric phase gate consists of an initial (dynamic) $\pi/2$ pulse that creates the spin coherence in the $xy$-plane, the adiabatic control sequence, and a (dynamic) $\pi$ pulse that refocuses the random fluctuations of the environment [Fig.~\ref{ad_demonstration}(a) inset]. The spin echo is detected and its phase is determined by quadrature detection, from which the phase acquired by the electron spin during the adiabatic phase gate can be deduced.

Figure~\ref{ad_demonstration}(a) shows the phase of the electron spin $\gamma$ measured after an adiabatic phase gate that is designed to apply a geometric phase $\phi$ to the spin, and the corresponding error defined as $\varepsilon_\gamma = \gamma - \phi$. The experimental data follows the theoretical relation $\gamma=\phi$ very well over a broad range of $\phi$, from 0 to $20\pi$, which verifies that we have successfully implemented the adiabatic geometric phase gate to the electron spins. The intensity of the spin echo is also plotted against $\phi$ in Fig.~\ref{ad_demonstration}(b) to illustrate the performance of the phase gate. The echo intensities are normalized to a Hahn echo for the same time delay. The fact that the echo intensity at $\phi=0$ is less than 1 indicates that the adiabatic process is not perfect, and implies only partial adiabatic following of the whole spin ensemble. This is partly due to the off-resonance error of the spins, however, since the ESR linewidth of the spins is narrower than the bandwidth of the adiabatic control sequence, the failure of adiabatic following is more generally due to the non-adiabaticity of the phase gate operation. This also explains why the echo intensity decreases for greater $\phi$: for a fixed duration of the adiabatic sequence; a greater $\phi$ implies a faster phase variation in the $xy$ plane during $t_d<t<t_d+t_s$, hence a less adiabatic operation. In addition, because of the off-resonance error and inhomogeneities in the microwave field, different spin packets do not follow exactly the same path during the adiabatic operation, and by the end of the evolution they will exhibit a spread in the final phase. While the geometric phase $\gamma$ is a measurement of the mean phase of all the spins, the spin echo intensity reflects the variance of phases of between different spins. The reduction of the echo intensity can thus be attributed to the loss of phase coherence between the spins. It is also responsible for the increasing uncertainty in $\phi$ [i.e., $\vert \epsilon_r \vert$ in Fig.~\ref{ad_demonstration}(a)], since the signal-to-noise ratio of the measurement suffers from the loss of echo intensity.

The effect of varying $t_s$, which determines the sweep rate of $\phi$ in the $xy$ plane, is shown in Figs.~\ref{ad_demonstration}(c) and (d). In accordance with the adiabatic condition, the phases measured with shorter $t_s$ contain larger errors. Furthermore, the echo intensity is reduced for shorter $t_s$ implying higher non-adiabaticity. The dips in the echo intensity traces for $t_s=3~\mu$s and $4~\mu$s are attributed to the non-adiabaticity of the phase gate rather than noise from the spectrometer, as we have observed similar features in simulations where the only imperfection introduced is the microwave field inhomogeneity. 

\begin{figure}[t]
\begin{center}
\includegraphics[width=3.5in]{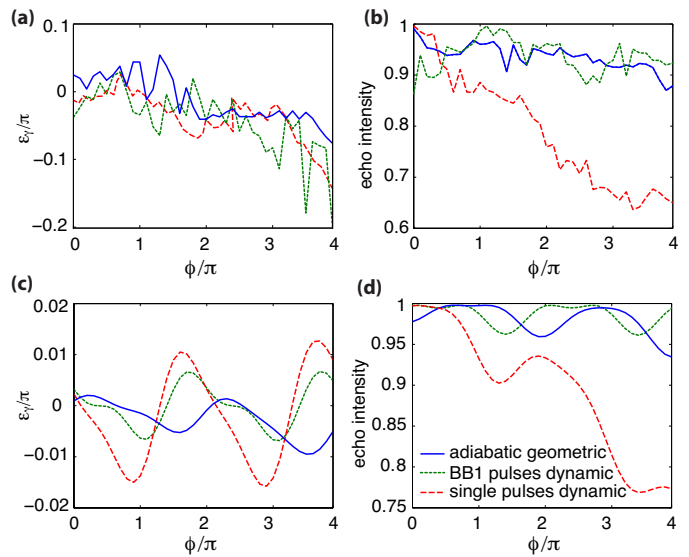}
\caption{(Color online) \textbf{(a)}: Measured phase error and \textbf{(b)}: echo intensity for the adiabatic geometric phase gate, and dynamic phase gates using single microwave and BB1 composite pulses. All intensities are normalized to unity to enable a more convenient comparison. The parameters used for the adiabatic sequence of the experiment are the same as in Fig.~\ref{ad_demonstration}. \textbf{(c)} and \textbf{(d)}: Simulation using the parameters $\Omega_0=1.4$~MHz, $\Delta_0=6$~MHz, $t_d=2~\mu$s, and $t_s=4~\mu$s.}
\label{exp_comparison1}
\end{center}
\end{figure}

We proceed by comparing the adiabatic phase gate to non-adiabatic phase gates based on single microwave and BB1 composite pulses. The gates are studied in the range of $[0,4\pi]$ since both the BB1 pulse operation and the non-adiabatic geometric gate have a natural phase limit of $4\pi$. The pulse sequences for the non-adiabatic gates are as described above for the simulations. Limited by the output level of the signal generator, the range of the linear amplification of the spectrometer, and our solid-state amplifier, the maximum amplitude of the microwave field we can apply is about 0.25~G, corresponding to the length of a $\pi$ pulse $\tau_\pi=700$~ns. In this case, the BB1 pulse sequence for a dynamic $\pi$ phase gate is $14\times\tau_\pi=9.8~\mu$s, therefore we choose $t_s=6~\mu$s and $t_d=2~\mu$s for the adiabatic sequence so that its total length $\tau=10~\mu$s is comparable to the BB1 pulse sequence. 

The measured phase error and echo intensity shown in Figs.~\ref{exp_comparison1}(a) and \ref{exp_comparison1}(b), respectively, demonstrate that the adiabatic phase gate outperforms its dynamic counterpart using single microwave pulses, while its performance is comparable to a dynamic gate with BB1 composite pulses. This is consistent with our simulations [Figs.~\ref{exp_comparison1}(c) and \ref{exp_comparison1}(d)] which also predict that the adiabatic geometric phase gate is more robust than the simple dynamic gate under $B_1$ inhomogeneity. 
However, we note that the measured error of Fig.~\ref{exp_comparison1}(a) is larger than the simulated one of Fig.~\ref{exp_comparison1}(c), explaining why the oscillatory features of the simulation are not visible in the experimental data. By contrast, the simulation in Fig.~\ref{exp_comparison1}(d) reproduces the main features in the echo intensity of the dynamic phase gate using single pulses (red, dashed trace), which indicates that the reduction in the echo intensity is essentially due to the $B_1$ inhomogeneity.
We have  performed a similar comparison of the adiabatic and the non-adiabatic geometric phase gate, which gives the same qualitative results, i.e.,~BB1 pulses need to be employed for the non-adiabatic implementation to obtain comparable performance. For the simulations, we assumed 10\% inhomogeneity in $B_1$ and no off-resonance effect. 

\section{Discussion}

In conclusion, we have introduced and demonstrated single-qubit geometric phase gates using adiabatic control of electron spins. Experiments and simulations showed that our adiabatic geometric phase gate is remarkably robust against inhomogeneities in the microwave field. 

A previous theoretical study has suggested that a slow adiabatic process is more exposed to environmental decoherence, mitigating its  advantage over non-adiabatic operations~\cite{Nazir2002}.  However, in our experiment the gate times for the BB1 and adiabatic phase gate are similar, and the fidelity of the adiabatic geometric phase gate is still limited by other imperfections of the equipment such as the small amplitude of the microwave field and phase imprecision. 

For the current experimental setup its performance is comparable to the geometric phase gate using composite non-adiabatic pulses such as BB1 pulses. However, the adiabatic phase gate is expected to be advantageous given a more inhomogeneous microwave field, such as may arise in coplanar resonators~\cite{Malissa2012}, or at higher microwave amplitudes, for example, achieved using a higher Q-value resonator~\cite{Borneman2012}.

\section{Acknowledgements}

This research was supported by the EPSRC through the Materials World Network (EP/I035536/1), the European Research Council under the European Community's Seventh Framework Programme (FP7/2007-2013) / ERC Grant Agreement No. 279781, and the National Research Foundation and Ministry of Education, Singapore. H.W. is supported by KCWong Education Foundation, M.M. thanks the Academy of Finland  (Grants No. 135794 and 251748) and Emil Aaltonen Foundation for financial support, and J.J.L.M. is supported by the Royal Society.
\\ ~ \\
\appendix
\section{Microwave Pulse Sequence}
\label{appendix}

The driving microwave field for implementing the adiabatic geometric phase gate of the electron spin is not uniquely determined, since any field that evolves adiabatically along the curve $\mathcal{C}_2$ in  Fig.~\ref{sequence}(a) will induce a geometric phase $\phi$ to the spin. Generally, our proposed sequence can  be divided into four sections with respective durations $t_d, t_s/2, t_s/2, t_d$: (i) the frequency sweep from off-resonance to resonance, (ii) the phase sweep in the $xy$-plane from 0 to $\phi/2$ before the $\pi$ phase shift, (iii) the phase sweep from $\phi/2$ to $\phi$ after the $\pi$ phase shift, and (iv) the frequency sweep from resonance to off-resonance. The total duration of the sequence is then $\tau = 2 t_d + t_s$, and any driving field must satisfy the following conditions:
\begin{subequations}	\label{conditions}
\begin{align}
~&\Omega_x(0)=\Omega_y(0)=0, ~\Omega_x(\tau)=\Omega_y(\tau)=0;  \nonumber \\
~&\Delta(t)=0, {\rm ~for} ~ t_d \le t \le t_d+t_s;   \nonumber \\
~&\Delta(t) =  \dot{\Delta}(t) =0  {\rm ~at}~t=t_d {\rm ~and}~t=t_d+t_s,   \nonumber
\end{align}
\end{subequations}
where the last condition is required to for achieving adiabaticity. In addition, the first derivative of the Hamiltonian with respect to time must be continuous, and the sequence needs to be symmetric about its midpoint for the dynamic phase to fully cancel.

For our study we employed a driving microwave field with the following frequency profile [{\it cf.} Fig.~\ref{sequence}(b)]

\begin{eqnarray}
\omega(t) = \left\{ \begin{array}{ll}
\omega_0 + \frac{\Delta_0}{2} \left[ \cos \left( \pi \frac{t}{t_d} \right)+1 \right] & 0 \le t < t_d,\\
\omega_0 & t_d \le t < \tau - t_d,  \\
\omega_0 + \frac{\Delta_0}{2} \left[ \cos \left( \pi\frac{t-\tau+t_d}{t_d} \right) -1 \right]  &
\tau-t_d \le t \le \tau. \nonumber
\end{array} \right.
\end{eqnarray}
where $\Delta_0=\omega(0) - \omega_0$ denotes the initial detuning at $t=0$.

In a rotating frame with the frequency of the driving microwave field, the amplitude and phase of the microwave drive shown in Fig.~\ref{sequence}(b) were given by
\begin{eqnarray}
\Omega(t) = \left\{ \begin{array}{ll}
\frac{\Omega_0}{2} \left[1-\cos \left( \pi \frac{t}{t_d} \right) \right] & 0 \le t < t_d,\\
\Omega_0 & t_d \le t < \tau - t_d,  \\
\frac{\Omega_0}{2} \left[ 1+\cos \left( \pi \frac{t-\tau+t_d}{t_d} \right) \right] & \tau - t_d  \le t \le \tau.
\end{array} \right. \nonumber
\end{eqnarray}
and 
\begin{eqnarray}
\varphi(t) = \left\{ \begin{array}{ll}
0 & 0 \le t < t_d,\\
\frac{\phi}{4} \left[ 1- \cos \left( 2 \pi \frac{t-t_d}{t_s} \right) \right] & t_d \le t < \tau/2,  \\
\frac{\phi}{4} \left[ 3- \cos \left( 2 \pi \frac{t-\tau/2}{t_s} \right) \right] +\pi & \tau/2 \le t <
\tau-t_d,  \\
0 & \tau-t_d \le t \le \tau.
\end{array} \right. \nonumber
\end{eqnarray}
The $\Omega_{x,y}(t)$ in Fig.~\ref{sequence}(c) correspond to the actual in-phase (I) and quadrature (Q) signals applied to the low-frequency inputs of the utilized (IQ) mixer and are obtained by taking a combined result of all the traces of Fig.~\ref{sequence}(b).


\begin{thebibliography}{99}

\bibitem{Nielsen2000}
M. A. Nielsen and I. L. Chuang, \emph{Quantum Computing and Quantum Information} 
(Cambridge University Press, Cambridge, 2000).

\bibitem{Schweiger2001}
A. Schweiger and G. Jeschke,
\emph{Principles of Pulse Electron Paramagnetic Resonance} (Oxford University Press, 2001).

\bibitem{Morton2005a} 
J. J. L. Morton, A. M. Tyryshkin, A. Ardavan, K. Porfyrakis \emph{et al.},
Phys. Rev. A, \textbf{71}, 012332 (2005).

\bibitem{Wimperis1994} 
S. Wimperis, J. Magn. Reson., Ser. A, \textbf{109}, 221 (1994).

\bibitem{Morton2005} 
J. J. L. Morton, A. M. Tyryshkin, A. Ardavan, K. Porfyrakis \emph{et al.},
Phys. Rev. Lett., \textbf{95}, 200501 (2005).

\bibitem{Ryan2010} 
C.~A. Ryan, J.~S. Hodges and D.~G. Cory, 
Phys. Rev. Lett., \textbf{105}, 200402 (2010).

\bibitem{Souza2011} A.~M. Souza, G.~A. \'Alvarez and D. Suter, 
Phys. Rev. Lett., \textbf{106}, 240501 (2011).

\bibitem{Berry1984}
M. V. Berry, Proc. R. Soc. London, Ser. A, \textbf{392}, 45 (1984).

\bibitem{Wilczek1984}
F. Wilczek and A. Zee, Phys. Rev. Lett., \textbf{52}, 2111 (1984).

\bibitem{Aharonov1987}
Y. Aharonov and J. Anandan, Phys. Rev. Lett., \textbf{58}, 1593 (1987).

\bibitem{Zanardi1999} 
P. Zanardi and M. Rasetti, Phys. Lett. A \textbf{264} 94, (1999).

\bibitem{Zhu2003}
S.-L. Zhu and Z. D. Wang, Phys. Rev. Lett., \textbf{91}, 187902 (2003).

\bibitem{Ekert2000}
A. Ekert, M. Ericsson, P. Hayden, H. Inamori \emph{et al.},
J. Mod. Opt., \textbf{47}, 2501 (2000).

\bibitem{Zhu2005}
S.-L. Zhu, P. Zanardi, Phys. Rev. A \textbf{72} 020301 (2005). 

\bibitem{Solinas2010}
P. Solinas, J.-M. Pirkkalainen, M. M\"ott\"onen, Phys. Rev. A \textbf{82}, 052304 (2010).

\bibitem{Oreshkov2009}
O. Oreshkov, T. A. Brun and D. A. Lidar, 
Phys. Rev. Lett., \textbf{102}, 070502 (2009).

\bibitem{Suter1987}
D. Suter, G. C. Chingas, R. A. Harris and A. Pines, 
Mol. Phys., \textbf{61}, 1327 (1987).

\bibitem{Jones2000}
J. A. Jones, V. Vedral, A. Ekert and G. Castagnoli, 
Nature, \textbf{403}, 869 (2000).

\bibitem{Du2003}
J. Du, P. Zou, M. Shi, L.~C. Kwek \emph{et al.},
Phys. Rev. Lett., \textbf{91}, 100403 (2003).

\bibitem{Leibfried2003}
D. Leibfried, B. DeMarco, V. Meyer, D. Lucas \emph{et al.},
Nature, \textbf{422}, 412 (2003).

\bibitem{Leek2007}
P.~J. Leek, J.~M. Fink, A. Blais, R. Bianchetti \emph{et al.}, 
Science, \textbf{318}, 1889 (2007).

\bibitem{Mottonen2008}
M. M\"ott\"onen, J.~J. Vartiainen and J.~P. Pekola, 
Phys. Rev. Lett., \textbf{100}, 177201 (2008).

\bibitem{DeChiara2003}
G. DeChiara and G.~M. Palma,
Phys. Rev. Lett., \textbf{91}, 090404 (2003). 

\bibitem{Filipp2009}
S. Filipp, J. Klepp, Y. Hasegawa, C. Plonka-Spehr \emph{et al.}, 
Phys. Rev. Lett., \textbf{102}, 030404 (2009).

\bibitem{Nazir2002}
A. Nazir, T.~P. Spiller and W.~J. Munro, 
Phys. Rev. A, \textbf{65}, 042303 (2002).

\bibitem{Malissa2012} 
H. Malissa, D. I. Schuster, A. M. Tyryshkin, A. A. Houck \emph{et al.}, 
Rev. Sci. Instrum. \textbf{84}, 025116 (2013).

\bibitem{Borneman2012} 
T. W. Borneman and D. G. Cory, 
J. Magn. Reson. \textbf{225}, 120 (2012).

\end{thebibliography}
\end{document}